\documentclass[sigconf,natbib]{acmart}
\usepackage{tcolorbox}
\usepackage[frozencache, cachedir=minted-cache]{minted}
\usepackage[english]{babel}
\usepackage{multirow}
\usepackage{colortbl}
\usepackage{enumitem}
\usepackage{caption}
\usepackage{tabularx}
\usepackage{array}
\newcolumntype{x}{>{\raggedright\arraybackslash}X}
\newcolumntype{R}{>{\raggedleft\arraybackslash}X}
\usepackage{multicol}
\usepackage{xcolor}

\definecolor{summaryGray}{RGB}{242, 242, 242}
\definecolor{borderSummaryGray}{RGB}{64, 64, 64}

\newcommand{\mybox}[1]{
\vspace{0.5em}
\noindent
{ 
  \setlength{\fboxrule}{0.2pt}
  \fcolorbox{borderSummaryGray}{summaryGray}{\parbox{.975\columnwidth}{#1}}
} 
\vspace{0.5em}
}

\setlist[itemize,1]{label=*, leftmargin=*} 

\AtBeginDocument{%
  }

\setcitestyle{numbers}

\begin{document}

\title{Decoding the Configuration of AI Coding Agents: Insights from Claude Code Projects}





\author{Hélio Victor F. Santos, Vitor Costa, João Eduardo Montandon and Marco Tulio Valente}
\affiliation{%
  \institution{Universidade Federal de Minas Gerais}
  \city{Belo Horizonte}
  \state{Minas Gerais}
  \country{Brazil}
}
\email{{helio.santos, vitorcosta, joao, mtov}@dcc.ufmg.br}

\begin{abstract}
  Agentic code assistants are a new generation of AI systems capable of performing end-to-end software engineering tasks. While these systems promise unprecedented productivity gains, their behavior and effectiveness depend heavily on configuration files that define architectural constraints, coding practices, and tool usage policies. However, little is known about the structure and content of these configuration artifacts. This paper presents an empirical study of the configuration ecosystem of Claude Code, one of the most widely used agentic coding systems. We collected and analyzed 328 configuration files from public Claude Code projects to identify (i) the software engineering concerns and practices they specify and (ii) how these concerns co-occur within individual files. The results highlight the importance of defining a wide range of concerns and practices in agent configuration files, with particular emphasis on specifying the architecture the agent should follow.

\end{abstract}

\maketitle

\section{Introduction}

Agentic code assistants are autonomous AI systems that independently execute complex software development workflows. Emerging in 2024 with tools like Claude Code (Anthropic), Codex (OpenAI), and Jules (Google), these agents mark a fundamental shift from code completion tools---such as the early versions of GitHub Copilot---to systems capable of performing end-to-end software engineering tasks. Rather than requiring constant human guidance, they autonomously plan solutions, generate and refactor code, debug issues, run tests, query documentation, and perform version control operations, including creating commits and submitting pull requests.

Although agentic code assistants are built for autonomy, they should be configured with rules that guide their behavior and help to improve their performance and effectiveness. In software projects, such rules may define the system architecture, enforce coding standards, or specify which tools---such as linters, test frameworks, or build systems---the agent can rely on. Essentially, the goal
is to allow autonomy while keeping development aligned with project  guidelines, tools, and best practices.

However, despite the growing popularity of agentic code assistants, little is known about how their configuration files are structured and what they reveal about agent behavior and capabilities. To address this gap, this paper explores the configuration ecosystem of Claude Code, one of the most advanced agentic coding systems. We collected and analyzed a dataset of 328 configuration files from publicly available and popular Claude Code projects. Using this dataset, we seek to answer two research questions:

\begin{itemize}[label=\textbullet]
  \item {\em RQ1: What concerns and practices are specified in agent configuration files?} Our goal is to help developers who are using---or planning to use---agentic systems configure their code agents effectively. To achieve this goal, we believe it is important to identify the software engineering aspects and practices most frequently specified in the configuration files of agentic coding systems. In particular, we focus on the configuration files of a widely used system, Claude Code.

  \item \textit{RQ2: In addition to textual rules, do configuration files include other elements such as code examples, links, and diagrams?} The goal is to determine whether these elements---which have specific tags in markdwon---are also used by developers when writing configuration files for code agents.

  \item \textit{RQ3: What are the most common patterns of Claude.md files?}
        With this RQ we intend to provide a list of typical agent configuration files so developers can choose which setup best fits their needs.
        For this, we analyzed which concerns and practices co-occur in the configuration files analyzed in RQ1, and reveal the most common patterns.
\end{itemize}

The remainder of this paper is organized as follows. Section \ref{sec:claude-code} introduces the Claude Code tool, and Section \ref{sec:study-design} describes our study design, including data collection, manual analysis, and pattern analysis. Sections \ref{sec:rq1}, \ref{sec:rq2}, and \ref{sec:rq3} present the results for each research question. Finally, Section \ref{sec:threats-to-validity} discusses threats to validity, Section \ref{sec:related-work} reviews related work, and Section \ref{sec:conclusion} concludes the paper.

\section{Claude Code}\label{sec:claude-code}

Claude Code is a command line tool for agentic coding created by Anthropic, and designed to assist developers in their coding workflows.\footnote{ https://www.anthropic.com/news/claude-3-7-sonnet}
Among its features, the tool can read and understand the codebase of a project, integrate with the local development environment to automate tasks, and manage Git workflows.

To make best use of Claude Code, developers can provide specific instructions about their project and how the agent should interact with it.
These instructions are defined in a configuration file named {\tt Claude.md} written in markdown format.
For each request, Claude first reads the content of this file looking for information that can help it at fulfilling the request.
These files contain a wide range of information, including general source code style guides, project's directory structure, and a list of commands to build, test, and run the project.
Figure \ref{fig:claude-md} presents a simple {\tt Claude.md} file.\footnote{This example was extracted and adapted from Claude Code's official documentation: \url{https://www.anthropic.com/engineering/claude-code-best-practices}}
As we can see, the file defines the bash commands Claude Code can execute, the code style used in the project, and general workflow guidelines, such as the recommendation to run single tests whenever possible, rather than the entire test suite, for performance reasons.

\begin{figure}[!h]
  \begin{minted}[
    frame=lines,
    framesep=2mm,
    breaklines,
    autogobble,
    fontsize=\footnotesize
    ]{markdown}
    # CLAUDE.md
    This file provides guidance to Claude Code
    
    ## Bash commands
    - npm run build: Build the project
    - npm run typecheck: Run the typechecker
    
    ## Code style
    - Use ES modules (import/export) syntax, not CommonJS (require)
    - Destructure imports when possible (eg. import { foo } from 'bar')
    
    ## Workflow
    - Be sure to typecheck when you're done making a series of code changes
    - Prefer running single tests, and not the whole test suite, for performance
  \end{minted}
  \caption{Example of Claude.md file.}
  \label{fig:claude-md}
\end{figure}

\section{Study Design} \label{sec:study-design}

In this section, we present the design and methodology used in the study, including the construction of the dataset and the procedures we used to answer the two RQs proposed in the research.

\subsection{Data Collection}\label{sec:data-collection}
This process was carefully planned and then divided into three steps, as detailed below.

\noindentparagraph{Repositories Detection.}
We first collected an initial list of GitHub repositories which are using Claude Code between August 28, 2025 and August 30, 2025. To this end, we combined distinct queries performed on the GitHub Search API to find repositories containing the file {\tt Claude.md}. This initial search returned 4,724 repositories.

\noindentparagraph{Repositories Selection.}
We selected repositories relevant to our study, i.e.,~projects that are popular, active and represent real-world applications.
We removed repositories based on the following criteria:

\begin{enumerate}
  \item \textit{Popularity.} We filtered out repositories with less than 100 stars, which is a common proxy to select popular projects in empirical studies~\cite{Borges2018}.
  \item \textit{Language.} We excluded non-English repositories, by checking its name and description.
  \item \textit{Actual Applications.} We removed repositories that are not real-world applications, e.g.,~awesome lists, tutorials, or configuration examples; this filtering step was performed manually, by analyzing the repository description and README.
\end{enumerate}

From the remaining ones, we selected the top-100 most popular repositories, i.e.,~those with the highest number of stars.
Overall, the projects are implemented in 23 programming languages, with JavaScript/TypeScript (35 projects), Python (16 projects), and Go (9 projects) being the most common ones. All projects are fairly popular (median of 950 stars) and active (median of 488 commits).
They are also long-lived, with a median age of 58 months since their first commit.
Some examples of selected projects include: \texttt{modelcontextprotocol/python-sdk} (the official Python SDK for implementing MCP servers and clients)\footnote{\url{https://github.com/modelcontextprotocol/python-sdk}}, \texttt{oven-sh/bun}\footnote{\url{https://github.com/oven-sh/bun}} (a popular JavaScript runtime environment), and \texttt{callstack/react-native\-testing-library} (React Native testing library).\footnote{https://github.com/callstack/react-native-testing-library} Finally, we fetched the \texttt{CLAUDE.md} files from the selected repositories.

\noindentparagraph{Fetching CLAUDE.md Files.} Finally, we fetched the \texttt{CLAUDE.md} files from the selected repositories.
During this process, we detected 45 files functioning as memory banks, i.e., files that only pointed to other markdown files within their respective projects.
In such cases, we retrieved the contents of the referenced files instead.
At the end of this process, we obtained a total of 328 \texttt{CLAUDE.md} files.

\subsection{Manual Analysis}\label{sec:manual-analysis}

We parsed all 328 {\tt Claude.md} files in our dataset to extract their section titles.
In markdown, section titles are represented by hashtags (\#), where the number of hashtags indicates the section level (e.g.,~\# for level 1, \#\# for level 2, and so on).
To answer the proposed RQs, we focused on level 2 sections---those starting with \#\#---since level 1 sections are typically used for the main title of the file (e.g., {\tt Claude.md}, as presented in Figure~\ref{fig:claude-md}). Thus, since Claude Code configuration files are written in Markdown, we leveraged their section structure to identify the categories of rules specified in these files, rather than analyzing each line individually.

Table \ref{tab:size-claude} shows the distribution of the number of level-2 headings in the dataset. As we can see, the files have a median of seven level-2 headings. However, there is also one file with 213 headings. Interestingly, we found that 36 files have no level-2 sections at all. After a manual analysis, we concluded that these files describe practices and rules directly at level 1 (\# tag). Thus, in such cases, we decided to analyze the content of level-1 headings instead.

\begin{table}[!ht]
  \caption{Size of Claude.md files (measured in number of \#\# sections).}
  \label{tab:size-claude}

\begin{tabular*}{\columnwidth}{l @{\extracolsep{\fill}} ccccc}
    \toprule
                  & \textbf{Min} & \textbf{Q1} & \textbf{Median} & \textbf{Q3} & \textbf{Max} \\ \midrule
    \#\# sections & 0          & 4.0         & 7.0             & 10.0        & 213       \\
    \bottomrule
  \end{tabular*}
\end{table}

After following the previous procedures, we extracted 2,492 section titles.
These sections were then manually analyzed by one of the authors, who grouped them into semantically related categories according to the software engineering concerns and practices they refer to.
For example, the sections listed below were grouped into a single practice called \textbf{Testing}: \textit{Running Tests}, \textit{Test Organization}, \textit{Testing}, \textit{Testing Approach}, \textit{Testing Considerations}, \textit{Testing Guidelines}, \textit{Testing Patterns}, \textit{Testing Purpose}, \textit{Testing Requirements}, \textit{Testing Strategy}, and \textit{Testing Structure}. Next, a meeting was held with two other authors, who inspected the proposed classification and confirmed it. The final classification, with the original section titles (\#\# tags), the proposed classification, and a link to the respective {\tt Claude.md} file is available in our replication package (link at the end of this manuscript).

\subsection{Pattern Analysis}


We used the FP-Max algorithm to detect which concerns and practices are mentioned together~\cite{grahne2003efficiently}.
FP-Max is a variant of the Apriori algorithm~\cite{agrawal1996fast} that focuses on finding maximal frequent itemsets, i.e.,~the largest independent sets of items that frequently co-occur in a dataset.
We relied on this algorithm to detect independent combinations of concerns and practices that are frequently described in the same {\tt Claude.md} files, thus representing possible file patterns.
We represent each {\tt Claude.md} file as a transaction; a tuple (or set) of the concerns and practices it contains.
Each tuple is then provided to the FP-Max algorithm to identify frequently co-occurring items.
We used the FP-Max implementation provided by the MLxtend library,\footnote{\url{https://rasbt.github.io/mlxtend/}} configured with 0.15 minimum support.

\section{Most Common Concerns and Practices (RQ1)}\label{sec:rq1}

\begin{figure}[htbp]
  \centering
  \includegraphics[width=\columnwidth]{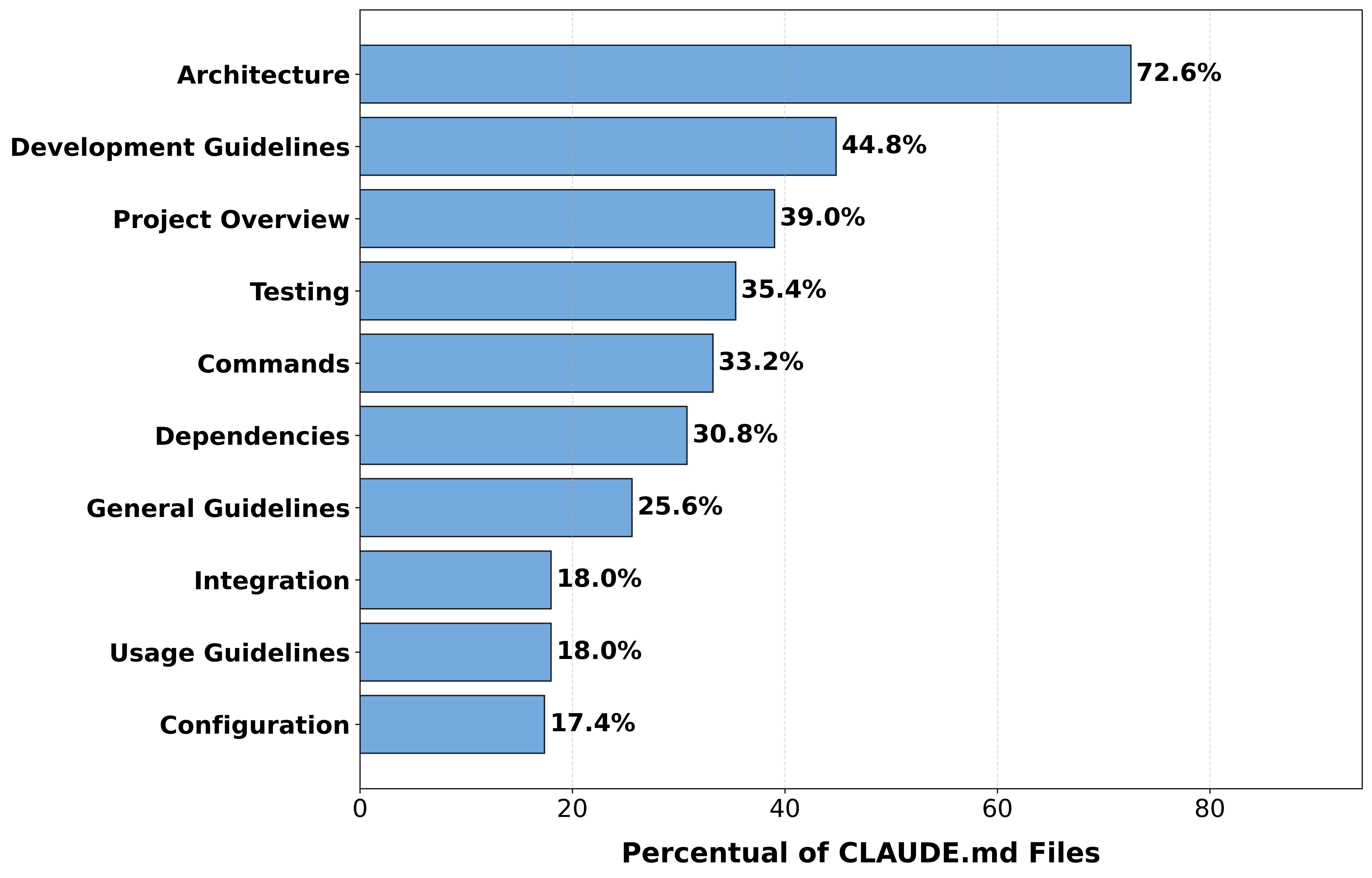}
  \caption{Concerns and practices addressed in agent config.~files}
  \label{fig:h2_titles}
\end{figure}

Figure \ref{fig:h2_titles} shows the main concerns and practices defined in coding agent configuration files. As we can see, these concerns and practices are related to the architecture of the software to be generated with the support of AI agents (found in 238 Claude.md files from our dataset, 72.6\%), followed by general development guidelines (147 Claude.md files, 44.8\%), project overview (128 Claude.md files, 39\%), and testing guidelines (116 Claude.md files, 35.4\%) and commands (109 Claude.md files, 33.2\%). Other recurring themes include dependencies (30.8\%), general project guidelines (25.6\%), integration and usage guidelines, each with 59 Claude.md files (18\%), and configuration, each with 57 Claude.md files (17.4\%). In the following, we will detail these common concerns and practices and provide real examples of recommendations that can be defined within them.\vspace{5pt}

\noindent\textbf{Software Architecture:} Essentially, these sections define the software architecture the coding agent should follow. For example, Figure \ref{fig:architecture-ex1} shows an example of such guidelines  in \texttt{evstack/ev-node}.\footnote{\url{https://github.com/evstack/ev-node/blob/main/CLAUDE.md}} The guidelines define the main packages of the architecture, key interfaces, and modular design principles (e.g., {\em each component has an interface in the core package}).\vspace{5pt}

\begin{figure}[htbp]
  \begin{minted}[
    frame=lines,
    framesep=2mm,
    breaklines,
    autogobble,
    fontsize=\footnotesize
    ]{markdown}
    ## Code Architecture
    
    ### Core Package Structure
    The project uses a zero-dependency core package pattern:
    - **core/** - Contains only interfaces and types, no external dependencies
    - **block/** - Block management, creation, validation, and synchronization
    - **p2p/** - Networking layer built on libp2p
    - **sequencing/** - Modular sequencer implementations
    - **testapp/** - Reference implementation for testing

    ### Key Interfaces
    - **Executor** (core/executor.go) - Handles state transitions
    - **Sequencer** (core/sequencer.go) - Orders transactions
    - **DA** (core/da.go) - Data availability layer abstraction

    ### Modular Design
    - Each component has an interface in the core package
    - Implementations are in separate packages
    - Components are wired together via dependency injection
    - Multiple go.mod files enable modular builds
  \end{minted}
    


  \caption{Architecture guidelines (evstack/ev-node)}
  \label{fig:architecture-ex1}
\end{figure}

\noindent\textbf{Development Guidelines:} This section aims to guide the AI agent on low-level development issues.
Figure \ref{fig:development-guidelines-ex1} shows an example from the \texttt{PrefectHQ/marvin} project.\footnote{\url{https://github.com/PrefectHQ/marvin/blob/main/CLAUDE.md
  }} It includes rules for typing, installing dependencies, navigating on the code, etc.\vspace{5pt}

\begin{figure}[htbp]
  \begin{minted}[
    frame=lines,
    framesep=2mm,
    breaklines,
    autogobble,
    fontsize=\footnotesize
    ]{markdown}
    ## Development Guidelines

    ### Type Hints
    - Use `X | Y` instead of `Union[X, Y]`
    - Use builtins like `list`, `dict` instead of `typing.List`, `typing.Dict`
    - Use `T | None` instead of `Optional`

    ### Dependencies & Running
    - Use `uv` for dependency management and script execution
    - Install deps: `uv sync` or `uv sync --extra foo`
    Run scripts: `uv run some/script.py` or `uv run --with pandas script.py`
    Testing: `uv run pytest` or `uv run pytest -n3` for parallel
    
    ### Finding Things
    - Use `rg` for searching, not grep
    - Use `ls` and `tree` for navigation

    ### Linter Philosophy
    - Empirically understand by running code
    - Linter tells basic truths but may be orthogonal to goals
  \end{minted}

  \caption{Development guidelines (PrefectHQ/marvin)}
  \label{fig:development-guidelines-ex1}
\end{figure}

\noindent\textbf{Commands:} This section documents commands that Claude Code can use in specific scenarios.
For example, for executing tests, formatting code, and performing type checking.

\noindent\textbf{Project Overview:}
This section is commonly used to provide Claude Code with context about the system through a general description. This helps the AI agent understand the objectives and the problem domain, resulting in more accurate code aligned with the system's needs. An example is shown in Figure \ref{fig:project-overview-ex2}, which clarifies the purpose and goals of the React Native Testing Library (RNTL).

\begin{figure}[htbp]
  \begin{minted}[
    frame=lines,
    framesep=2mm,
    breaklines,
    autogobble,
    fontsize=\footnotesize
    ]{markdown}
    ## Project Overview
    
    This is the **React Native Testing Library (RNTL)** - a comprehensive testing solution for React Native applications that provides React Native runtime simulation on top of `react-test-renderer`. The library encourages better testing practices by focusing on testing behavior rather than implementation details.
  \end{minted}
  \caption{Project overview (callstack/react-native-testing-library)}
  \label{fig:project-overview-ex2}
\end{figure}

\noindent\textbf{Testing:} This section defines testing concerns and practices, such as in the example in Figure~\ref{fig:testing-ex1}, where the configuration specifies the types of tests that the agents should create (unit, integration tests, and mocks) and the test commands.

\begin{figure}[htbp]
  \begin{minted}[
    frame=lines,
    framesep=2mm,
    breaklines,
    autogobble,
    fontsize=\footnotesize
    ]{markdown}
    ## Testing
    **Test Structure:**
    - Unit tests for core utilities
    - Integration tests for dbt operations
    - Mock data for warehouse connections
    
    **Test Commands:**
    ```bash
    jest                     # Run all tests
    jest --watch            # Watch mode
    jest src/dbt/models.test.ts  # Specific test file
    ```
    \end{minted}

  \caption{Example of testing guidelines (lightdash/lightdash)}
  \label{fig:testing-ex1}
\end{figure}

\mybox{{\em Summary:} Our results highlight the importance of defining a wide range of concerns and practices in agent configuration files, with emphasis on specifying the architecture the agent should follow when generating the application (as found in 72.6\% of the analyzed {\tt Claude.md} files).
}

\section{Usage of Code Examples and Links  (RQ2)}\label{sec:rq2}

Table \ref{tab:usage-code-example-links} presents the occurrence of source code examples and links in the analyzed {\tt Claude.md} files. As shown, the highest percentage of code examples appears in the Development Guidelines category (17.68\% of the instances classified in this category).

\begin{table}[!ht]
  \caption{Usage of Code Example and Links}
  \label{tab:usage-code-example-links}

  \begin{tabularx}{\columnwidth}{Xrr}
    \toprule
                           & \textbf{Code Examples} & \textbf{Links} \\ \midrule
    Architecture           & 10.98\%                & 1.83\%         \\
    Development Guidelines & 17.68\%                & 0.61\%         \\
    Testing                & 15.24\%                & 0.0\%          \\
    Commands               & 15.55\%                & 0.3\%          \\
    \bottomrule
  \end{tabularx}
\end{table}

An example is provided in Figure \ref{fig:code-example}, which includes a code snippet demonstrating how to configure a custom database table using Supabase. 


\begin{figure}[!h]
  \begin{minted}[
    frame=lines,
    framesep=2mm,
    breaklines,
    autogobble,
    fontsize=\footnotesize
    ]{markdown}
    ## Common Patterns
    
    ### Custom Table Configuration
    ```typescript
    const vectorStore = new SupabaseVectorStore({
      supabaseUrl: process.env.SUPABASE_URL,
      supabaseKey: process.env.SUPABASE_KEY,
      table: "custom_table_name",
    });
    ```
  \end{minted}
  \caption{Config.~file with code example (run-llama/LlamaIndexTS)}
  \label{fig:code-example}
\end{figure}

Conversely, links were most frequently found in the Architecture category, though in only 1.83\% of the files. Figure \ref{fig:example-link} illustrates an example containing a link to a file documenting design patterns and code conventions. Finally, we also searched for diagrams in the files—specifically UML diagrams written in the syntax of the  Mermaid\footnote{\url{https://mermaid.js.org}} tool, but
 found only two instances.\vspace{5pt}

\begin{figure}[htbp]
  \begin{minted}[
    frame=lines,
    framesep=2mm,
    breaklines,
    autogobble,
    fontsize=\footnotesize
    ]{markdown}
    ## Design patterns and Code conventions
    
    Please see the dedicated "[Design patterns and Code conventions](design-patterns-and-conventions.md)" page.
  \end{minted}
  \caption{Config.~file with a link to a project's page (grafana/mimir)}
  \label{fig:example-link}
\end{figure}

\mybox{{\em Summary:} In the analyzed configuration files, there is a non-negligible presence of code examples, especially in sections that describe Development Guidelines (17.68\%).
}

\section{Patterns of Configuration Files (RQ3)}\label{sec:rq3}

Table \ref{tab:frequent-sections-set} presents the top-5 most common \texttt{Claude.md} patterns detected by the FP-Max algorithm.
These patterns encompass combinations of two to three concerns and practices.
The most common one includes practices on \textit{Architecture}, \textit{Dependencies}, and \textit{Project Overview}, appearing in 21.6\% of the files;
the fifth pattern appears in 17.7\% of the files, and contains \textit{Architecture} and \textit{Integration} rules.
\textit{Architecture} stands out as being the unique practice described in all top-5 patterns.
\textit{Development Guidelines} and \textit{Project Overview} show up twice, each.
The remaining ones---\textit{Dependencies}, \textit{General Guidelines}, \textit{Testing}, and \textit{Integration}---appear only once.\vspace{5pt}

\begin{table}[!ht]
  \caption{Top-5 Most Popular \texttt{Claude.md} Patterns.}
  \label{tab:frequent-sections-set}
  \small
  \begin{tabularx}{\columnwidth}{Xr}
    \toprule
    \textbf{Practices \& Concerns}                         & \multicolumn{1}{l}{\textbf{\% Files}} \\
    \midrule
    Architecture, Dependencies, Project Overview           & 21.6                                  \\
    \textit{Example: Repository Structure, Key Dependencies, Project Overview}                     \\ \midrule
    Architecture, General Guidelines                       & 20.1                                  \\
    \textit{Example: Repository Structure, Specialized Agent Usage Guidelines}                     \\ \midrule
    Architecture, Development Guidelines, Project Overview & 19.8                                  \\
    \textit{Example: Package Structure, Component Development Guidelines, Overview}                \\ \midrule
    Architecture, Development Guidelines, Testing          & 18.9                                  \\
    \textit{Example: Code Architecture, Development Workflow, Testing}                             \\ \midrule
    Architecture, Integration                              & 17.7                                  \\
    \textit{Example: Architecture Overview, External Integrations}                                 \\
    \bottomrule
  \end{tabularx}
\end{table}

\mybox{{\em Summary:} The most common pattern of {\tt Claude.md} file includes rules about Architecture, Dependencies, and Project Overview (21.6\%). Moreover, Architecture is a key element in the identified patterns, appearing in the top-5 ones.}

\section{Threats To Validity} \label{sec:threats-to-validity}

Even though this is an initial study, we highlight two threats to validity. First, the classification of concerns and practices specified in the {\tt Claude.md} files was performed by only one author, although it was reviewed and confirmed by two additional authors. Second, the field of AI agents is highly dynamic, meaning that our results may evolve with advances in language models and agent-based tools. However, we believe such changes will not be drastic, since the current results reflect classical Software Engineering concerns and practices, such as architecture and testing.

\section{Related Work}\label{sec:related-work}
Large Language Models (LLMs) are transforming software engineering practices. Recent research describes this new era as ``Software Engineering 3.0'', where AI plays a key role in increasing productivity and reducing the cognitive load of developers~\cite{hassan2024ainativesoftwareengineeringse}.
As developers are increasingly relying on such agents to perform software engineering tasks~\cite{yang2024sweagentagentcomputerinterfacesenable, Lambiase_2024, Dakhel2023copilot}, investigating good practices for configuring and working with these tools becomes essential.

Research in this area has primarily focused on evaluating the effectiveness of these agents in solving software problems, using benchmarks like SWE-bench to measure performance~\cite{jimenez2024swebench}. The Agentless approach, for example proposed a three-step workflow to solve bugs and other issues using LLMs~\cite{xia2024agentlessdemystifyingllmbasedsoftware}.
Other studies investigated the practical use of these tools and their impact on software projects.
\citet{Tufano2024unveiling} and \citet{Watanabe2024chatgpt} have investigated how developers use LLM-based chatbots, such as ChatGPT, for various software engineering tasks.
\citet{Silva2024} analyzed the effectiveness of ChatGPT for detecting code smells in Java Projects.
\citet{watanabe2025useagenticcodingempirical} analyzed 567 pull requests generated by Claude Code on GitHub and highlighted the importance of incorporating project-specific rules into the agent's instructions.
Our work proposes to investigate the guidelines that humans provide to configure software agents, thus offering a new lens on the need of collaboration between developers and AI agents.

\section{Conclusion} \label{sec:conclusion}

This paper analyzed 328 Claude Code configuration files to understand how developers configure AI coding agents. The results show that these files capture key software engineering practices, especially those related to architectural concerns. As future work, we plan to analyze discussions about coding agent configurations in pull requests; study the evolution and modifications made to such files; and implement a tool to recommend best practices for writing agent configuration files.\vspace{5pt}

\noindent{\bf Replication Data:} The data and results of this research are available at: \url{https://doi.org/10.5281/zenodo.17388127}

\begin{acks}
  This research was supported by grants from CNPq and FAPEMIG.
\end{acks}

\bibliography{main}
\end{document}